\documentclass{jfm}

\usepackage{graphicx}
\usepackage{newtxtext}
\usepackage{newtxmath}
\usepackage{placeins}
\usepackage{natbib}
\usepackage{hyperref}
\hypersetup{
    colorlinks = true,
    urlcolor   = blue,
    citecolor  = black,
}

\newcommand{\RomanNumeralCaps}[1]
\linenumbers

\title{An analytical model of momentum availability for predicting large wind farm power}

\author{Andrew Kirby\aff{1}
  \corresp{\email{andrew.kirby@trinity.ox.ac.uk}}, Thomas D. Dunstan\aff{2}, \and Takafumi Nishino\aff{1}}

\affiliation{\aff{1}Department of Engineering Science, University of Oxford, Parks Road, Oxford OX1 3PJ, UK
\aff{2}Met Office, FitzRoy Road, Exeter EX1 3PB, UK}

\begin{document}
\maketitle

\begin{abstract}Turbine-wake and farm-atmosphere interactions influence wind farm power production. For large offshore farms, the farm-atmosphere interaction is usually the more significant effect. This study proposes an analytical model of the `momentum availability factor' to \textcolor{blue}{predict the impact of} farm-atmosphere interactions. It models the effects of net advection, pressure gradient forcing and turbulent \textcolor{blue}{entrainment, using} steady quasi-1D flow assumptions. Turbulent entrainment is modelled by assuming self-similar vertical shear stress profiles. We used the model with the `two-scale momentum theory' to predict the power of large finite-sized farms. The model compared well with existing results of large-eddy simulations (LES) of finite wind farms in conventionally neutral boundary layers. The model captured most of the effects of \textcolor{blue}{atmospheric boundary layer (ABL) height} on farm performance by considering the undisturbed vertical shear stress profile of the ABL \textcolor{blue}{as an input}. \textcolor{blue}{In particular, the} model  predicted the power of staggered wind farms with a typical error of 5\% or less. The developed model provides a novel way of \textcolor{blue}{instantly} predicting the power of large wind farms, including the farm blockage effects. \textcolor{blue}{A further simplification of the model to analytically predict the 'wind extractability factor' is also presented.} This study provides a \textcolor{blue}{novel} framework for modelling \textcolor{blue}{farm-atmosphere interactions}. Future studies can use the framework to better model large wind farms.  
\end{abstract}

\begin{keywords}
Authors should not enter keywords on the manuscript, as these must be chosen by the author during the online submission process and will then be added during the typesetting process (see \href{https://www.cambridge.org/core/journals/journal-of-fluid-mechanics/information/list-of-keywords}{Keyword PDF} for the full list).  Other classifications will be added at the same time.
\end{keywords}

{\bf MSC Codes }  {\it(Optional)} Please enter your MSC Codes here

\section{Introduction}

\par Wind energy is a key technology for the renewable energy transition. To meet future energy demands, wind energy capacity will need to increase rapidly and individual farms will likely become larger \citep{Veers2022}. A key aspect for designing wind farms is predicting their power output. However, it is difficult to model the aerodynamics of large wind farms because of the multi-scale nature of flows involved \citep{Porte-Agel2020}.

\par Traditionally, there are two main approaches to predicting wind farm performance at a low computational cost. For the first approach, semi-analytical `wake' models predict the velocity deficit in the wake behind a turbine \citep[e.g.][]{jensen1983,bastankhah2014}. To model an entire wind farm, individual wakes are superposed using different techniques \citep[e.g.][]{katic1986,zong2020}. Wake models are commonly used to optimise the layout of turbines in a farm. However, they do not consider the atmospheric response to wind farms and thus they perform poorly for extended wind farms \citep[e.g.][]{stevens2016}. The second approach uses `top-down' models \citep[e.g.][]{frandsen1992,frandsen2006,calaf2010}. Top-down models consider the response of an idealised atmospheric boundary layer (ABL) to an infinitely large wind farm.  They cannot however, predict the impact of turbine layout \textcolor{blue}{(placement of turbines within a farm)} on farm performance. Recent studies have coupled the wake and top-down models together \citep[e.g.][]{stevens2016,starke2021}. This approach however, is still limited by the limitations of the constituent models, e.g., idealised ABL profiles and wake superposition methods.

\par \textcolor{blue}{More recently, models have been developed to predict the interaction between wind farms and the atmosphere. \citet{Meneveau2012} extended top-down models to finite wind farms by considering the development of an internal boundary layer. \citet{Luzzatto-Fegiz2018} modelled the impact of entrainment on farm performance by using a three layer model. This model was extended to finite wind farms by \citet{Bempedelis2023}. \citet{Smith2010} predicted the impact of gravity waves on wind farm performance by solving a two layer model using fast fourier transforms. This approach was later extended to a three layer model \citep{Allaerts2018,Allaerts2019}. The model was further extended to include the impact of vertically varying free-atmospheres \citep{Devesse2022}. All of these approaches require solving differential equations numerically to predict wind farm performance.}

\par \citet{Nishino2020} developed the `two-scale momentum theory' to \textcolor{blue}{better understand the power generation mechanism} of large wind farms. This splits the multi-scale problem of wind farm aerodynamics into `internal' turbine/array-scale and `external' farm/atmospheric-scale sub-problems. The sub-problems are coupled together using the conservation of momentum. As farms become larger, one of the grand challenges facing the wind energy community is to understand \textcolor{blue}{the impact of} farm-atmosphere interaction \citep{Veers2019}. \citet{Kirby2022} \textcolor{blue}{confirmed this by performing} large-eddy simulations (LES) of 50 different periodic turbine layouts. They introduced the concepts of `turbine-scale losses' due to turbine-wake interactions and `farm-scale losses' caused by the atmospheric response to the whole farm. For large offshore farms, the farm-scale losses were found to be typically more than twice as large as the turbine-scale losses. This highlights the importance of modelling farm-scale flows to predict the performance of future large farms.

\par Unlike most large wind farm models, the two-scale momentum theory does not assume any specific profiles of the ABL (such as a logarithmic law), allowing for the external sub-problem to be modelled in various manners. For the external modelling, it is crucial to accurately predict how the momentum available to the farm site increases due to the presence of the turbines, i.e., the momentum availability. Recent studies \citep{Patel2021,legris2023} used `twin' numerical weather prediction (NWP) simulations to calculate the momentum availability and thus predict how farm-scale losses (including the so-called farm blockage losses) change with atmospheric conditions. However, this approach would be too computationally expensive for wind farm optimisation unless a sufficiently long set of twin NWP (such as those reported by \citet{Stratum2022}) has already been conducted for candidate wind farm sites.

\par In the present study we propose a simple analytical model to predict the momentum availability for large wind farms. This model, together with the two-scale momentum theory, allows us to predict the power of a large wind farm analytically. The model is derived using quasi-1D control volume analysis and assuming self-similar vertical shear stress profiles. In section \ref{two_scale_theory}, we summarise the two-scale momentum theory and the key wind farm parameters. Section \ref{M_model} presents the derivation of the new momentum availability factor model. In section \ref{les_comparison} we compare the predictions of farm power with existing finite wind farm LES. The model is \textcolor{blue}{further} discussed in section \ref{discussion} and concluding remarks are given in section \ref{conclusion}.

\section{Two-scale momentum theory}\label{two_scale_theory}

By considering the conservation of momentum for a control volume with and without a wind farm present, \citet{Nishino2020} derived the non-dimensional farm momentum (NDFM) equation:

\begin{equation}
    C_T^* \frac{\lambda}{C_{f0}} \beta^2 + \beta^\gamma = M
    \label{eqn:windfarmmomentum}
\end{equation}

\noindent where $\beta$ is the farm wind-speed reduction factor defined as $\beta\equiv U_F/U_{F0}$ (with $U_F$ defined as the average wind speed in the nominal farm-layer of height $H_F$, and $U_{F0}$ is the farm-layer-averaged speed without the presence of the turbines); $C_T^*$ is the (farm-averaged) `internal' turbine thrust coefficient defined as $C_T^*\equiv \sum_{i=1}^{n}T_i/\frac{1}{2}\rho U_F^2nA$ (where $T_i$ is the thrust of turbine $i$, $A$ is the rotor swept area and $n$ is the number of turbines in the farm); $\lambda$ is the array density defined as $\lambda\equiv nA/S_F$ (where $S_F$ is the wind farm area); $C_{f0}$ is the natural friction coefficient of the surface defined as $C_{f0}\equiv \tau_{w0} /\frac{1}{2}\rho U_{F0}^2$ (where $\tau_{w0}$ is the undisturbed bottom shear stress); $\gamma$ is the bottom friction exponent defined as $\gamma\equiv \log_\beta (\tau_w / \tau_{w0})$ (assumed to be 2 in this study); and $M$ is the momentum availability factor given by:

\begin{equation}
    M = \frac{X-C-\left[\frac{\partial p}{\partial x_F}\right] - \frac{\partial [\rho U]}{\partial t}}{X_0-C_0-\left[\frac{\partial p_0}{\partial x_{F0}}\right] - \frac{\partial [\rho U_0]}{\partial t}}
    \label{momentumavailability}
\end{equation}

\noindent where $U$ is the velocity in the hub-height wind direction (i.e. streamwise direction) $x_F$, $X$ represents the net streamwise momentum injection through the top and side boundaries of the control volume (due to advection and Reynolds stress), $C$ is the streamwise component of the Coriolis force averaged over the control volume, $\partial p / \partial{x_F}$ is the pressure gradient in the direction $x_F$ and the subscript $0$ refers to values without the turbines present. \textcolor{blue}{Any imbalance between the momentum supplied to the control volume and total bottom drag (i.e., turbine thrust and surface drag) will accelerate or decelerate the flow, giving the time derivative terms in equation \ref{momentumavailability} \citep{Nishino2020}.} In this study \textcolor{blue}{we ignore the Coriolis terms and time derivative terms (i.e., we assume stationary atmospheric conditions) and} use a fixed definition for the farm-layer height of $H_F=2.5H_{hub}$ where $H_{hub}$ is the turbine hub-height, following \citet{Kirby2022}.

\par The theoretical framework \citet{Nishino2020} used to derive equation \ref{eqn:windfarmmomentum} is known as the `two-scale momentum theory'. The left-hand side of equation \ref{eqn:windfarmmomentum} is expected to depend primarily on turbine-scale or `internal' conditions. This includes turbine layout and operating conditions, and the hub-height wind speed and direction. The right-hand side of equation \ref{eqn:windfarmmomentum} is assumed to depend on `external' farm-scale conditions.

\par The farm wind-speed reduction factor $\beta$ can be calculated using \ref{eqn:windfarmmomentum} for a set of $C_T^*$, $\gamma$ and $M$. Using $\beta$, the average turbine power coefficients can be calculated using

\begin{equation}
    C_p = \beta^3 C_p^*
    \label{cp}
\end{equation}

\noindent where $C_p$ is the (farm-averaged) turbine power coefficient defined as $C_p\equiv \sum_{i=1}^{n}P_i/\frac{1}{2}\rho U_{F0}^3nA$ ($P_i$ is power of turbine $i$ in the farm) and $C_p^*$ is the (farm-averaged) `internal' turbine power coefficient defined as $C_p^*\equiv \sum_{i=1}^{n}P_i/\frac{1}{2}\rho U_F^3nA$.

\section{Momentum availability factor model}\label{M_model}

The momentum availability factor $M$ describes the increase in momentum supplied to the farm site due to the presence of the turbines (equation \ref{eqn:M_decomp}). We introduce new variables $M_{F0}$ defined as the total net momentum flux into the farm control volume without the turbines present, and $M_F$ as the total net momentum flux with the turbines. $\Delta M_F$ is the change in momentum flux due to the presence of the turbines, defined as $\Delta M_F \equiv M_F - M_{F0}$. 

\par $\Delta M_F$ can be decomposed into the contributions from different physical mechanisms. For this study, we will consider only the contributions from net momentum advection, pressure gradient forcing (PGF) and turbulent entrainment (equation \ref{eqn:M_decomp}). Equation \ref{eqn:M_decomp} provides a new framework for modelling the impact of farm-scale flows.

\begin{eqnarray}
    M &=& \frac{M_F}{M_{F0}} = 1 + \frac{\Delta M_F}{M_{F0}} \nonumber\\
    && \mbox{}= 1 + \frac{\Delta M_{F,Advection}}{M_{F0}} + \frac{\Delta M_{F,PGF}}{M_{F0}} + \frac{\Delta  M_{F,Entrainment}}{M_{F0}}
    \label{eqn:M_decomp}
\end{eqnarray}

\par In this study we propose simple analytical models for the components of equation \ref{eqn:M_decomp}. For each component we consider a control volume of height $H_F$ around the farm (note that this control volume can be different from the control volume used to derive equation \ref{eqn:windfarmmomentum}). The height of the control volume can be chosen arbitrarily but choosing $H_F$ allows terms to be linked to farm wind-speed reduction factor $\beta$. We use a steady quasi-1D analysis for the advection and PGF terms.  The entrainment term is modelled by assuming self-similar vertical shear stress profiles (above the top turbine-tip). However, the modelling framework is not specific to the analytical models proposed in the next sections. Equation \ref{eqn:M_decomp} can be a starting point for future studies using more sophisticated approaches for each component. 

\subsection{Net momentum advection}

\begin{figure}
  \centerline{\includegraphics[width=\linewidth]{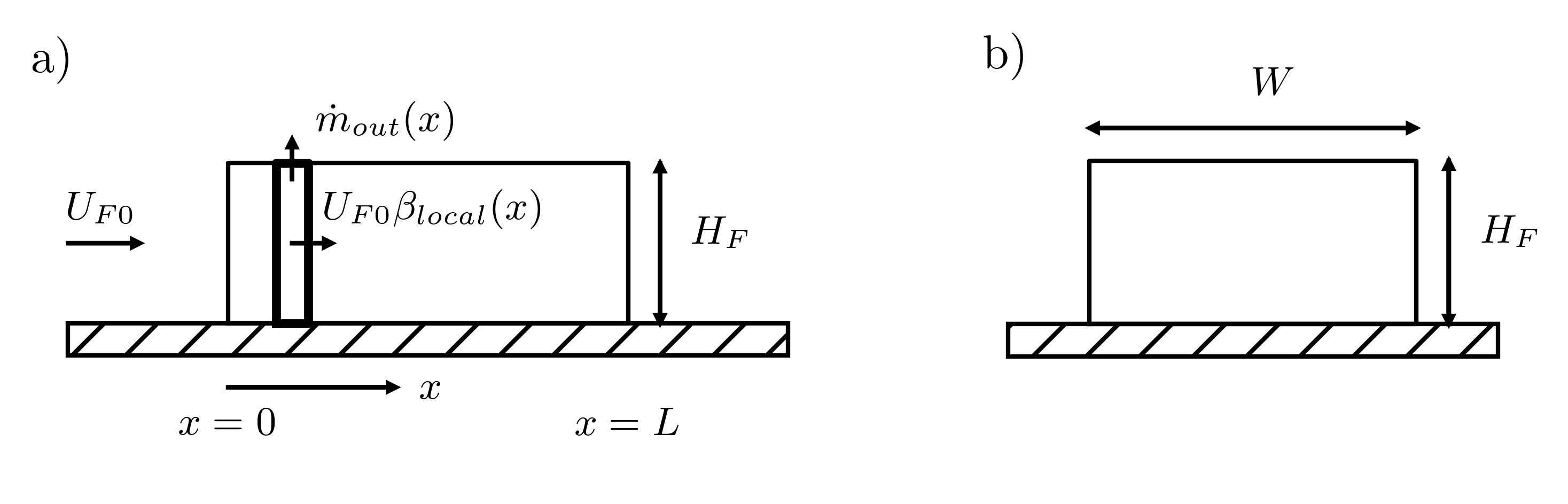}}
  \caption{Control volume analysis for net momentum advection calculation a) side view and b) front view.}
  \label{fig:net_advection}
\end{figure}

 Figure \ref{fig:net_advection} shows a rectangular wind farm control volume. $L$ is the length of the farm in the hub-height wind direction and $W$ is the farm width. Using a quasi-1D approach, we define the spanwise- and vertically-averaged wind speed throughout the control volume as $U_{F0}\beta_{local}(x)$. In our notation $x=0$ is \textcolor{blue}{at} the front of the farm and $x=L$ is \textcolor{blue}{at} the rear. Note that in figure \ref{fig:net_advection}a, $\beta_{local}(0)$ is not exactly equal to 1 since the wind speed decelerates upstream of the farm due to the farm blockage effect. Our proposed model of net momentum advection can therefore capture the impact of farm blockage.

\FloatBarrier

By considering the conservation of mass of an elemental control volume (shown by the bold box in figure \ref{fig:net_advection}), the mass flux out of the farm control volume at position $x$ can be expressed as

\begin{eqnarray}
    \dot{m}_{out}(x) = - \rho H_F W U_{F0} \frac{\mathrm{d}\beta_{local}(x)}{\mathrm{d}x}.
    \label{eqn:w_velocity}
\end{eqnarray}

\noindent Note that this could be mass flux out of the top or sides of the farm control volume. The momentum flux into the control volume at the front surface (i.e. $x=0$) is given by

\begin{equation}
  \Dot{m}_{in,front}U_{in,front} = \rho U_{F0}^2 \beta_{local}(0)^2H_FW
  \label{eqn:adv_front}
\end{equation}

\noindent noting that a positive value represents a net inflow of momentum. The net momentum flux through \textcolor{blue}{the} rear surface (i.e. $x=L$) is given by

\begin{equation}
  \Dot{m}_{in,rear}U_{in,rear} = - \rho U_{F0}^2 \beta_{local}(L)^2H_FW.
  \label{eqn:adv_rear}
\end{equation}

\noindent The momentum flux through the top surface is the integral of the mass flux and streamwise velocity at each position $x$, i.e.

\begin{eqnarray}
  \Dot{m}_{in,top}U_{in,top}  =  - \int_{0}^{L} \dot{m}_{out}(x) U_{F0} \beta_{local}(x) \,\mathrm{d}x.
  \label{eqn:top_int}
\end{eqnarray}

\noindent We substitute \ref{eqn:w_velocity} into \ref{eqn:top_int} and integrate to obtain

\begin{eqnarray}
 \Dot{m}_{in,top}U_{in,top} = \rho U_{F0}^2 H_FW \left[\frac{1}{2}\beta_{local}(L)^2 - \frac{1}{2}\beta_{local}(0)^2\right].
 \label{eqn:adv_top}
\end{eqnarray}

\noindent The total net momentum inflow into the control volume is given by the sum of equations \ref{eqn:adv_front}, \ref{eqn:adv_rear} and \ref{eqn:adv_top}, i.e., 

\begin{equation}
    \Delta M_{F,Advection} = \frac{1}{2} \rho U_{F0}^2H_FW\left[\beta_{local}(0)^2 - \beta_{local}(L)^2\right].
\end{equation}

\noindent This expression relates the change in momentum advection to the velocity at the front and rear of the farm.

\FloatBarrier

\subsection{Pressure gradient forcing}

\par
\begin{figure}
  \centerline{\includegraphics[width=\linewidth]{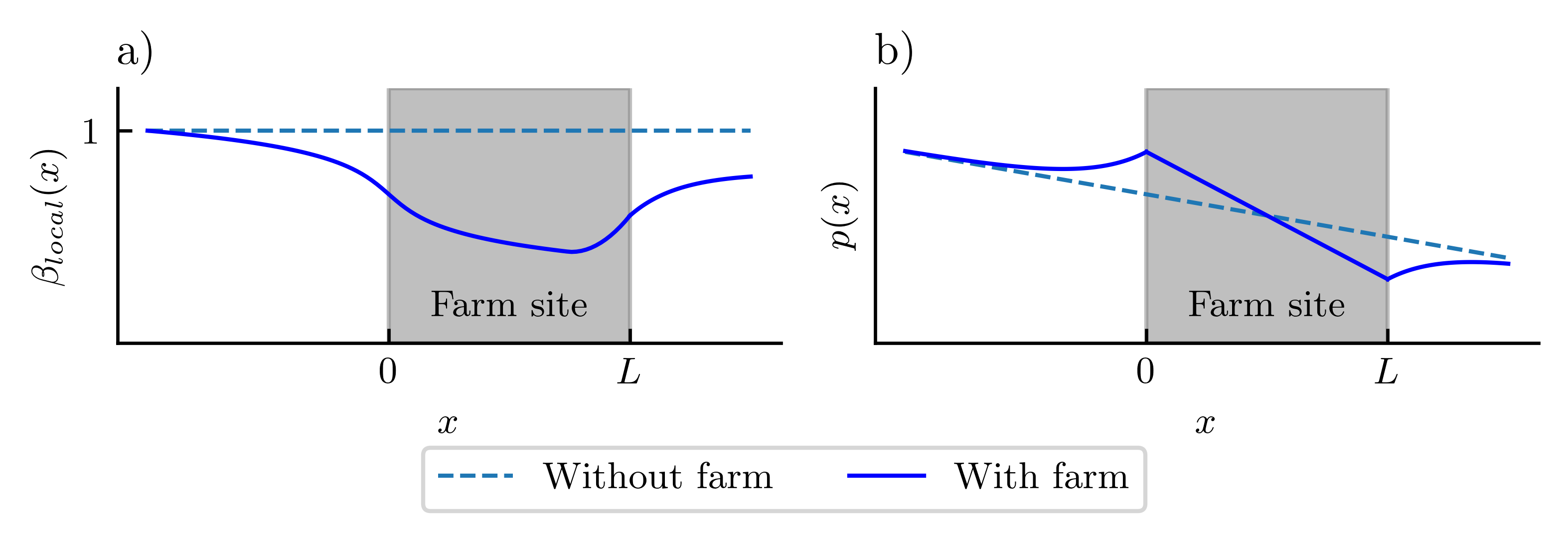}}
  \caption{Example variations of a) local farm wind-speed reduction factor $\beta_{local}(x)$ and b) pressure $p(x)$ with streamwise location.}
  \label{fig:PGF}
\end{figure}

 LES results in the literature have shown that large wind farms can induce additional pressure gradients across them \citep[e.g.,][]{Allaerts2017,Wu2017,Lanzilao2022}. When farms induce additional pressure gradients, a velocity reduction at the front of the farm is commonly observed. This velocity reduction at the front of the farm is commonly known as `wind-farm blockage'. Typically, a velocity increase and pressure reduction at the rear of the farm \textcolor{blue}{can also occur}. Figure \ref{fig:PGF} shows example streamwise variations of $\beta_{local}(x)$ and pressure.

\par In this section we propose a simple model for $\Delta M_{F,PGF}$. We apply Bernoulli's equation as an approximation to link the pressure increase to the velocity reduction at the front of the farm. We neglect changes in gravitational potential energy. It is assumed that changes in pressure are uniform up to the control volume height $H_F$.

\par \textcolor{blue}{Our aim here is to model an increased pressure difference across the farm $\Delta p$. The physical mechanism increasing the pressure difference is not directly considered (e.g. gravity waves) but is implicitly included in $\Delta p$. $\Delta p$ is composed of a pressure increase at the front surface of the farm control volume $\Delta p_{front}$ and pressure decrease at the rear surface $\Delta p_{rear}$.} 

\par \textcolor{blue}{The velocity reduction in front of the farm takes place over a distance of $L_{induction}$. Without the farm present, the background pressure force is balanced by the bottom shear stress over this region, i.e.,} 

\begin{equation}
    \textcolor{blue}{h_0\Delta p_0 = \frac{1}{2}\rho C_{f0}U_{F0}^2L_{induction}}
\end{equation}

\noindent \textcolor{blue}{where $h_0$ is the ABL height without the farm present. Therefore, the ratio of friction head loss to dynamic pressure is given by $L_{induction}C_{f0}/h_0$. Although $L_{induction}$ can be an order of magnitude larger than $h_0$, a typical value of $C_{f0}$ is 0.002 to 0.003. The dynamic pressure is therefore typically one to two orders of magnitude larger than the friction head loss and background pressure gradient. As such, as a first order approach we consider only pressure changes due to changes in dynamic pressure, justifying the use of Bernoulli's equation.}

\par Bernoulli's equation is applied on a streamline from far upstream to the front of the farm. The increase in pressure force on the front surface of the control volume is therefore given by:

\begin{equation}
    \Delta p_{front} H_FW= \frac{1}{2} \rho U_{F0}^2H_FW\left[1-\beta_{local}(0)^2\right].
\end{equation}

\noindent \textcolor{blue}{We then assume that $\Delta p_{rear} = \Delta p_{front}$ and therefore $\Delta p = 2\Delta p_{front}$. This is a strong assumption and in reality this would depend on the atmospheric conditions. The LES results of \citet{Wu2017} and \citet{Lanzilao2022} show approximately equal and opposite pressure changes at the front and rear of the farm, suggesting that this approximation is valid. However, the LES of \citet{Allaerts2017} and \citet{Lanzilao2023} suggest that this assumption is likely to only be valid for certain atmospheric stratifications.}

\noindent $\Delta M_{F,PGF}$ is then given by:

\begin{equation}
    \Delta M_{F,PGF} = \Delta p H_FW = \frac{1}{2} \rho U_{F0}^2H_FW\left[2 - 2\beta_{local}(0)^2\right].
\end{equation}

\noindent \textcolor{blue}{Combining} advection and PGF terms gives:

\begin{equation}
    \Delta M_{F,Advection} + \Delta M_{F,PGF}  = \frac{1}{2} \rho U_{F0}^2H_FW\left[2 - \beta_{local}(0)^2  - \beta_{local}(L)^2\right].
    \label{eqn:M_adv_pgf}
\end{equation}

\noindent Generally, the velocity at the front of the farm \textcolor{blue}{is} close to the undisturbed value. The velocity at the rear of the farm \textcolor{blue}{tends to} be close to the farm-averaged value. As a simple first-order approach, we assume that the \textcolor{blue}{small} velocity reduction at the front of the farm, $1-\beta_{local}(x)$, is equal to the \textcolor{blue}{small} increase at the rear (relative to the farm-averaged wind speed), $\beta_{local}(L)-\beta$. As such we can say that $\beta_{local}(0)^2 + \beta_{local}(L)^2 \approx 1 + \beta^2$. Applying this approximation to equation \ref{eqn:M_adv_pgf}:

\begin{equation}
    \Delta M_{F,Advection} + \Delta M_{F,PGF}  = \frac{1}{2} \rho U_{F0}^2H_FW\left(1 - \beta^2\right).
\end{equation}

\noindent \textcolor{blue}{To check the validity of this approximation, we used data from finite wind farm LES performed by \citet{Wu2017}. We take the hub-height wind speed normalised by the inflow value as a proxy for $\beta_{local}(x)$. Note that $H_F$ has been defined so that the hub-height wind speed is approximately equal to the farm-layer-averaged speed \citep{Kirby2022}. For the four farms simulated by \citet{Wu2017}, this assumption gives an overestimation on the order of 10\% (table \ref{tab:beta_error_pgf}).}

\begin{table}
  \begin{center}
\def~{\hphantom{0}}
  \begin{tabular}{c|cccc}
        & \multicolumn{2}{c}{$\Gamma=1K/km$}   & \multicolumn{2}{c}{$\Gamma=5K/km$}  \\[3pt]
          & Staggered & Aligned & Staggered & Aligned\\
        $\beta_{local}(0)^2+\beta_{local}(L)^2$ & 1.519 & 1.579 & 1.361 & 1.473\\
        $1 + \beta^2$ & 1.639 & 1.696 & 1.548 & 1.611\\
        Percentage error & 7.99\% & 7.40\% & 13.7\% & 9.37\%\\
  \end{tabular}
  \caption{Percentage errors of approximation $\beta_{local}(0)^2 + \beta_{local}(L)^2 \approx 1 + \beta^2$ using data from \citet{Wu2017}}
  \label{tab:beta_error_pgf}
  \end{center}
\end{table}

\noindent Dividing by the initial momentum supply $M_{F0}$ gives:

\begin{equation}
    \frac{\Delta M_{F,Advection}}{M_{F0}} +  \frac{\Delta M_{F,PGF}}{M_{F0}} = \frac{\frac{1}{2} \rho U_{F0}^2H_FW\left(1 - \beta^2\right)}{\tau_{w0}LW} = \frac{1}{C_{f0}}\frac{H_F}{L}\left(1-\beta^2\right).
    \label{eqn:adv_pgf}
\end{equation}

\noindent Although this is an approximation, equation \ref{eqn:adv_pgf} shows that when advection and PGF terms are combined, the dependence on farm inlet and outlet velocities disappears. Interestingly, this suggests that the impact of farm-scale flows may not directly depend on the wind speed at the front of the farm.

\FloatBarrier

\subsection{Turbulent entrainment}

\par Wind farms increase the turbulent mixing within the ABL. This increases the momentum entrainment into the farm \citep{stevens2017}. This mechanism supplies momentum to the control volume through the shear stress at the top surface. $\Delta M_{F,Entrainment}$ can be expressed as

\begin{equation}
    \frac{\Delta M_{F,Entrainment}}{M_{F0}} = \frac{\tau_tLW - \tau_{t0}LW}{\tau_{w0}LW} = \frac{\tau_t - \tau_{t0}}{\tau_{w0}}
    \label{eqn:entrainment}
\end{equation}

\noindent where $\tau_t$ is the shear stress at the top of the control volume.

\par The ABL can be stratified with different profiles of potential temperature which can change farm performance \citep{Porte-Agel2020}. In this study we consider conventionally neutral boundary layers (CNBLs) because many wind farm LES studies adopted ABLs with these profiles. CNBLs have self-similar vertical shear stress profiles (\citet{liu2021}). Therefore, if the surface stress and the boundary layer height are known, then the shear stress at any height can be determined. Note that the height is normalised by $h=h_{0.05}/(1-0.05^{2/3})$ where $h_{0.05}$ is the height where the shear stress is 5\% of the surface value.

\par
\begin{figure}
  \centerline{\includegraphics[width=\linewidth]{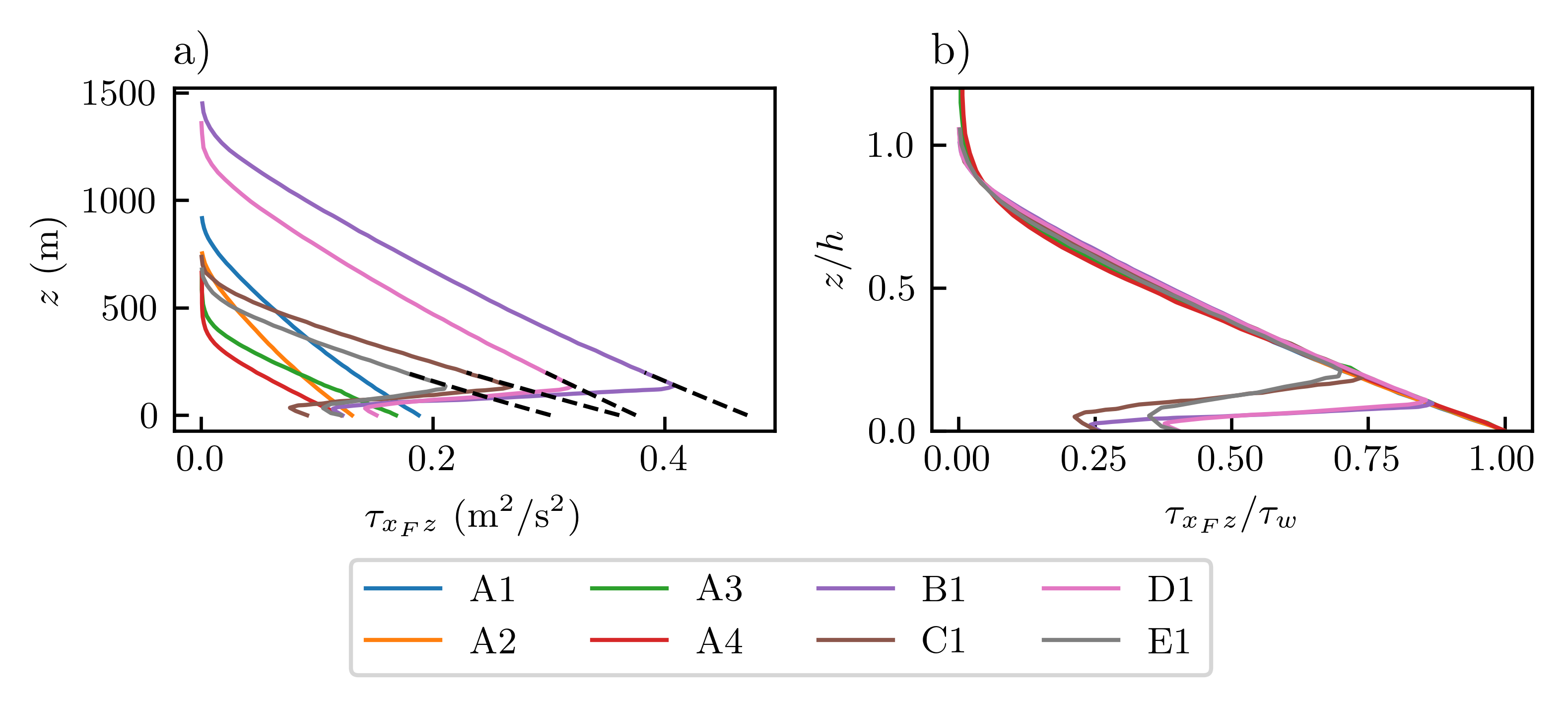}}
  \caption{a) Vertical profiles of shear stress from horizontally periodic LES with and without the farm present \citep{abkar2013} b) normalised vertical profiles.}
  \label{fig:stress_self_similar}
\end{figure}

 The self-similarity of shear stress profiles can \textcolor{blue}{also} be applied to large wind farms. \citet{abkar2013} performed horizontally periodic LES of CNBLs with and without turbines present. Figure \ref{fig:stress_self_similar}a shows the vertical profiles of the shear stress (note this is the stress in the hub-height wind direction, $x_F$). Above the turbine top tip ($126.5$m in \citet{abkar2013}) the vertical profiles have a similar shape. This suggests that the stress profiles above the turbines is equivalent to an `empty' CNBL with a higher surface stress. 

\par The black dotted lines in figure \ref{fig:stress_self_similar}a show the stress profiles above the turbines extrapolated to the surface (using a second order polynomial regression). This corresponds to the total bottom stress, i.e., surface shear stress and turbine thrust. When the stress profiles of \citet{abkar2013} are normalised by the total bottom stress and new ABL height, they fall onto the same curve (figure \ref{fig:stress_self_similar}b). This shows that the CNBLs containing wind farms also follow \textcolor{blue}{approximately} the same self-similar shear stress profile. Figure \ref{fig:stress_self_similar}b shows that wind farms increase the total bottom stress and boundary layer height of CNBLs.

\par
\begin{figure}
  \centerline{\includegraphics[width=0.5\linewidth]{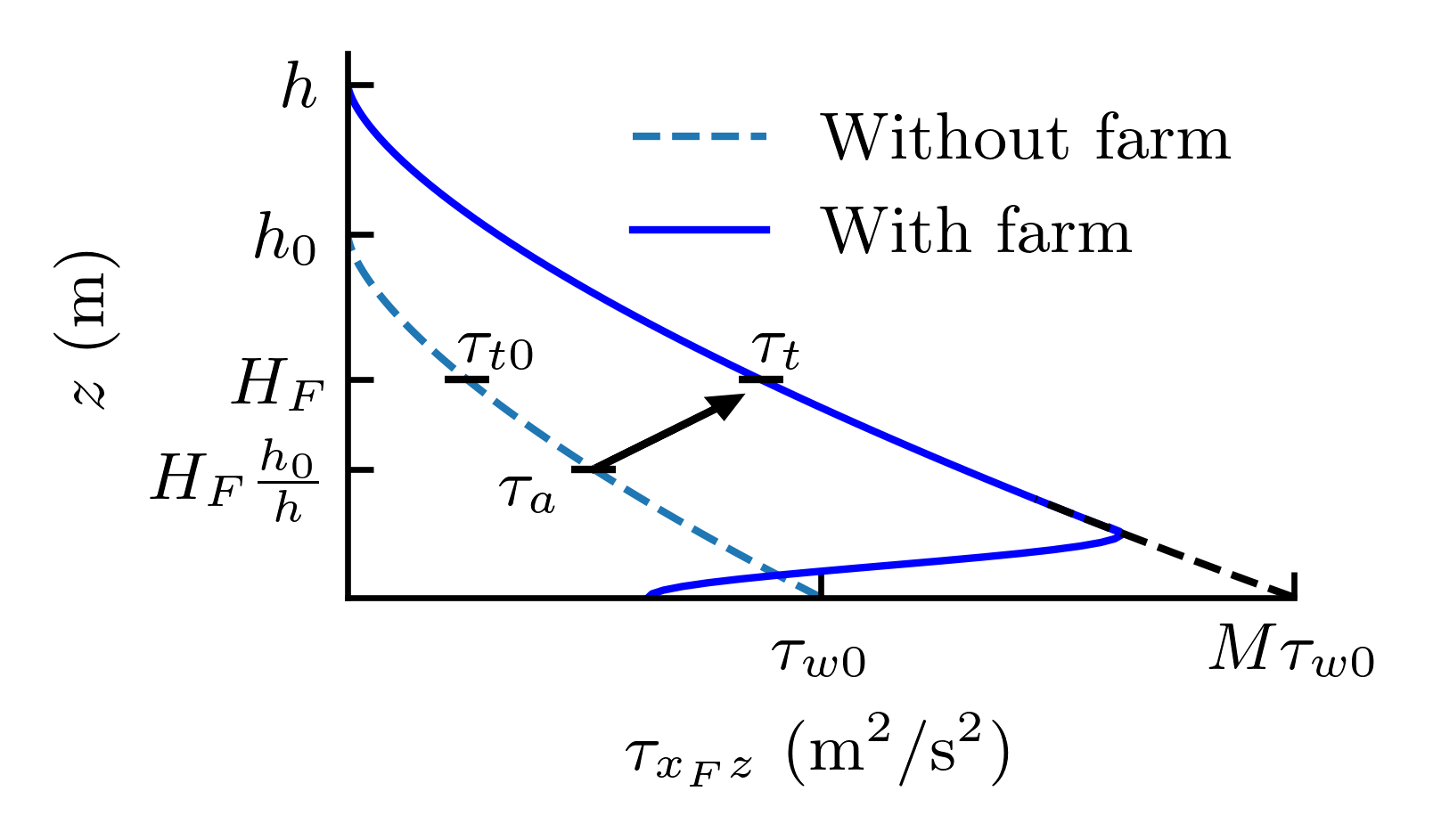}}
  \caption{Schematic of vertical shear stress profiles with and without the wind farm.}
  \label{fig:farm_stress}
\end{figure}

 The self-similarity of stress profiles can be used to predict momentum entrainment into large finite-sized wind farms. As a first order approach, we consider the vertical shear stress profile horizontally-averaged across the farm. Figure \ref{fig:farm_stress} shows a schematic of the shear stress profile with and without the farm present. Here, $h$ is the ABL height with the farm present averaged across the farm site. With the farm present, the shear stress is scaled by $M$ and the heights scaled by $h/h_0$.

\par To determine $\tau_t$, we need to calculate the undisturbed shear stress at a height of $H_Fh_0/h$ (denoted as $\tau_a$). This is found by linearly interpolating the stress between the top of the control volume and the surface, i.e.,

\begin{equation}
    \tau_a = \tau_{w0} - (\tau_{w0} - \tau_{t0}) \frac{h_0}{h}.
\end{equation}

\noindent Whilst the shear stress profile is not strictly linear, it is approximately linear away from the top of the ABL. $\tau_t$ can then be expressed in terms of $M$ and the undisturbed shear stress profile

\begin{eqnarray}
    \tau_t &=& M\tau_a = M \left(1 - \frac{h_0}{h}\right)\tau_{w0} + M\frac{h_0}{h}\tau_{t0}.
    \label{eqn:tau_top}
\end{eqnarray}

\par Equation \ref{eqn:tau_top} can be substituted into equation \ref{eqn:entrainment} to calculate $\Delta M_{F,Entrainment}$,

\begin{eqnarray}
        \frac{\Delta M_{F,Entrainment}}{M_{F0}} = \frac{(\tau_t-\tau_{t0})LW}{\tau_{w0}LW} &=& M + M\frac{h_0}{h} \left(\frac{\tau_{t0}}{\tau_{w0}}-1\right) - \frac{\tau_{t0}}{\tau_{w0}}
        \label{eqn:entrainment_final}
\end{eqnarray}

\par The different sources of momentum increase can be summed to calculate $M$. Subsituting equations \ref{eqn:adv_pgf} and \ref{eqn:entrainment_final} into \ref{eqn:M_decomp} we obtain

\begin{eqnarray}
M = 1 + \frac{1}{C_{f0}}\frac{H_F}{L}\left(1-\beta^2\right) + M + M\frac{h_0}{h} \left(\frac{\tau_{t0}}{\tau_{w0}}-1\right) - \frac{\tau_{t0}}{\tau_{w0}}
\end{eqnarray}

\noindent which can be rearranged to give

\begin{eqnarray}
M = \frac{1 + \frac{1}{C_{f0}}\frac{H_F}{L}\left(1-\beta^2\right)  - \frac{\tau_{t0}}{\tau_{w0}}}{\frac{h_0}{h} \left(1-\frac{\tau_{t0}}{\tau_{w0}}\right) }.
\label{eqn:M_general}
\end{eqnarray}

\par Using equation \ref{eqn:M_decomp} we can simply sum the different sources of momentum increase. The result is a single equation to model \textcolor{blue}{the impact of farm-scale flows on} farm performance. Equation \ref{eqn:M_general} is an approximate expression for $M$ as a function of $\beta$ and $h/h_0$. Different models could be used to predict $h/h_0$ in this formula for $M$. In the next section we present a simple first order model for $h/h_0$ as a function of $\beta$. 

\subsection{Boundary layer height increase}

\par
\begin{figure}
  \centerline{\includegraphics[width=\linewidth]{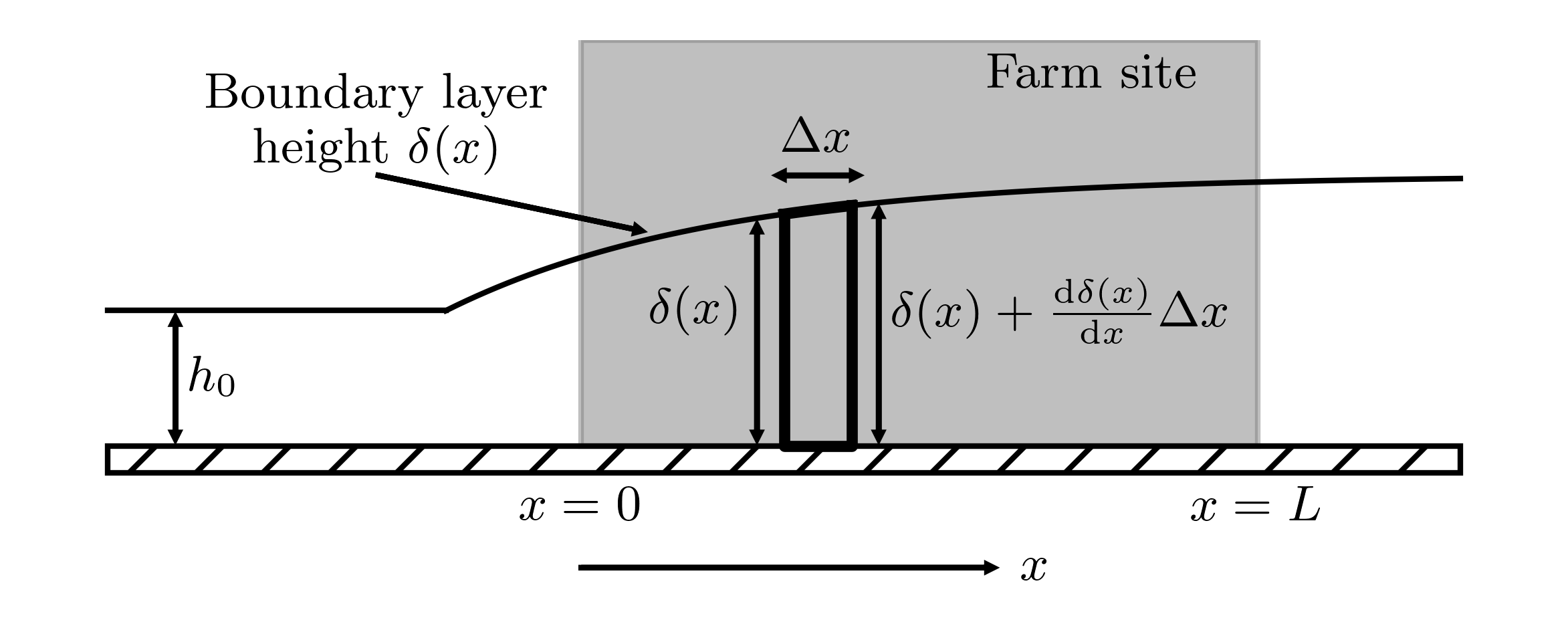}}
  \caption{Quasi-1D control volume analysis for ABL flow over a wind farm.}
  \label{fig:bl_height_increase}
\end{figure}

 In this section we present a simple model for increase in ABL height $h/h_0$ in response to a wind farm. We use a quasi-1D analysis of a boundary layer flow over a farm (see figure \ref{fig:bl_height_increase}). Here $\delta(x)$ is the ABL height at steamwise position $x$. We assume there is no mass flux between the boundary layer and free atmosphere above. The quasi-1D approach neglects the effect of flow \textcolor{blue}{bypassing} around the sides of the farm.

We consider the conservation of mass across a control volume encompassing a section of the ABL (in bold in figure \ref{fig:bl_height_increase}). The velocity averaged over the height of the ABL at position $x$ is denoted $U_A(x)$. We assume that the streamwise variation of $U_A(x)$ is the same as the streamwise variation of $\beta_{local}(x)$. This is strictly not true, but could give a reasonable first-order approximation of $U_A(x)$. Therefore $U_A(x)=U_{A0}\beta_{local}(x)$ where $U_{A0}$ is the velocity averaged throughout the undisturbed ABL. The mass flux in through the left hand side of the control volume is given by:

\begin{eqnarray}
\dot{m}_{in} = \rho U_{A0}\beta_{local}(x)\delta(x)W.
\end{eqnarray}

\noindent Taking the limit as $\Delta x$ approaches $0$, the mass flux out of the right hand side is given by:

\begin{eqnarray}
\dot{m}_{out} &=&\rho \left[U_{A0}\beta_{local}(x)+U_{A0}\frac{\mathrm{d}\beta_{local}(x)}{\mathrm{d}x}\mathrm{d}x\right]\left[\delta(x)+\frac{\mathrm{d}\delta(x)}{\mathrm{d}x}\mathrm{d}x\right] W\nonumber\\
\dot{m}_{out} &=& \rho U_{A0}\beta_{local}(x)\delta(x)W + \rho U_{A0}\beta_{local}(x)\frac{\mathrm{d}\delta(x)}{\mathrm{d}x}\mathrm{d}xW + \nonumber\\ && \rho U_{A0}\frac{\mathrm{d}\beta_{local}(x)}{\mathrm{d}x}\mathrm{d}x\delta(x)W
\end{eqnarray}

\noindent whilst neglecting second order terms. From the mass conservation, $\Dot{m}_{in}=\Dot{m}_{out}$,

\begin{eqnarray}
\rho U_{A0}\beta_{local}(x)\frac{\mathrm{d}\delta(x)}{\mathrm{d}x}W &=& -\rho U_{A0}\delta(x)\frac{\mathrm{d}\beta_{local}(x)}{\mathrm{d}x}W \nonumber\\
\frac{\mathrm{d}\delta(x)}{\delta(x)} &=& -\frac{\mathrm{d}\beta_{local}(x)}{\beta_{local}(x)}.
\label{eqn:bl_height_mass}
\end{eqnarray}

\noindent Both sides of equation \ref{eqn:bl_height_mass} can be integrated from far upstream to a given position, resulting in

\begin{eqnarray}
\delta(x) = \frac{h_0}{\beta_{local}(x)}
\end{eqnarray}

\noindent using the condition that at far upstream $\delta(x) =h_0$ and $\beta_{local}(x)=1$. To find $h$, $\delta(x)$ is averaged between $x=0$ and $x=L$ (as $h$ is the ABL height averaged across the wind farm site): 

\begin{eqnarray}
\frac{h}{h_0} = \frac{1}{L}\int_0^L \frac{\mathrm{d}x}{\beta_{local}(x)} \approx \frac{1}{\beta}
\label{eqn:bl_height_integral}
\end{eqnarray}

\noindent where the relationship between $\beta$ and $\beta_{local}(x)$ is given by $\beta = \int_0^L\beta_{local}(x)\mathrm{d}x$. If $\beta_{local}(x)$ is constant, the left hand side of equation \ref{eqn:bl_height_integral} is equal to $1/\beta$, which we assume to be generally true. In reality, the more $\beta_{local}(x)$ varies the less accurate this assumption becomes. If $\beta_{local}(x)$ is close to zero at any point the approximation in equation \ref{eqn:bl_height_integral} becomes less accurate. \textcolor{blue}{To check the validity of this assumption, we again use the LES data from \citet{Wu2017}.} For the four farms simulated by \citet{Wu2017}, the maximum error of this assumption was about 0.4\% (table \ref{tab:beta_error}). Therefore the approximation is reasonable for realistic profiles of $\beta_{local}(x)$.

\begin{table}
  \begin{center}
\def~{\hphantom{0}}
  \begin{tabular}{c|cccc}
        & \multicolumn{2}{c}{$\Gamma=1K/km$}   & \multicolumn{2}{c}{$\Gamma=5K/km$}  \\[3pt]
          & Staggered & Aligned & Staggered & Aligned\\
        $1/\beta$ & 1.250 & 1.200 & 1.349 & 1.281\\
        $\frac{1}{L}\int_0^L \frac{\mathrm{d}x}{\beta_{local}(x)}$ & 1.253 & 1.202 & 1.354 & 1.285\\
        Percentage error & 0.244\% & 0.160\% & 0.414\% & 0.248\%\\
  \end{tabular}
  \caption{Percentage errors of approximation in equation \ref{eqn:bl_height_integral} using data from \citet{Wu2017}}
  \label{tab:beta_error}
  \end{center}
\end{table}

\par \textcolor{blue}{The derivation of equation \ref{eqn:bl_height_integral} neglects horizontal flow deflections around the farm. This assumption is reasonable so long as the mass flux through the top surface of the farm control volume is much greater than through the sides. The mass flux through the top surface is given by $\rho wLW$ (where $w$ is the average vertical velocity through the top surface). The mass flux through the side surfaces is given by $2\rho vLH_F$ (where $v$ is the average lateral velocity at the side surfaces). If we suppose that $v$ and $w$ are of similar magnitudes, then neglecting horizontal deflections is reasonable so long as $W>>H_F$ (i.e. $2.5H_{hub}$).}

\par Substituting equation \ref{eqn:bl_height_integral} into equation \ref{eqn:M_general}, we get the following expression for $M$

\begin{eqnarray}
M &=& \frac{1 + \frac{1}{C_{f0}}\frac{H_F}{L}\left(1-\beta^2\right)  - \frac{\tau_{t0}}{\tau_{w0}}}{\beta \left(1-\frac{\tau_{t0}}{\tau_{w0}}\right)}.
\label{eqn:M_model}
\end{eqnarray}

\noindent Equation \ref{eqn:M_model} is a single \textcolor{blue}{algebraic} equation to predict the impact of farm-scale flows on farm performance. It includes the effects of net advection, PGF and turbulent entrainment. The only environmental parameter needed is the undisturbed shear stress profile. Equation \ref{eqn:M_model} can be used with the two-scale momentum theory to predict farm power. In section \ref{les_comparison} we compare the predictions of farm power output with the results from finite wind farm LES results reported in the literature.

\FloatBarrier

\section{Comparison with finite wind farm LES}\label{les_comparison}

\par The two-scale momentum theory presented in section \ref{two_scale_theory} can be used to predict wind farm power. For arrays of actuator-discs (or aerodynamically ideal turbines operating below the rated wind speed), \textcolor{blue}{$C_p^*=\alpha C_T^*$ where $\alpha \equiv U_T/U_F$ ($U_T$ is the streamwise velocity averaged over the rotor swept area). We can estimate $\alpha$ using the expresion $\alpha=\sqrt{C_T^*/C_T'}$ where $C_T'\equiv T_i/\frac{1}{2}\rho U_{T,i}^2 A$ is a turbine resistance coefficient describing the turbine operating conditions (note that this is strictly valid only for infinitely large regular
arrays of turbines where the thrust of each
individual turbine is identical to farm-averaged turbine thrust). The $C_p$ of an actuator disc is therefore given by} 

\begin{equation}
    \textcolor{blue}{C_p = \beta^3C_p^*=\beta^3 {C_T^*}^\frac{3}{2}{C_T'}^{-\frac{1}{2}}.}
    \label{cp_actuator_disc}
\end{equation}

\noindent \textcolor{blue}{Note that, when considering an isolated turbine, $C_T'$ is related to the classic turbine thrust coefficient by the relation $C_T'=C_T/(1-a)^2$ (where $a$ is the turbine induction factor).} $C_T^*$ generally depends on the turbine layout, but its upper limit may be predicted by using an analogy to the classical actuator-disc theory \citep{Nishino2016}, as 

\begin{equation}
    C_T^* = \frac{16C_T'}{(4+C_T')^2}.
    \label{eqn:ctstar_model}
\end{equation}

\noindent \textcolor{blue}{Note that equation \ref{eqn:ctstar_model}, like the classical actuator disc theory, is only valid for induction factors of up to 0.5.} Since this model \textcolor{blue}{predicts an upper limit of $C_T^*$ for a given $C_T'$}, it can be used to predict an upper limit to farm performance \citep{Kirby2022}. Using this model with $\beta=1$, the power coefficient of an isolated turbine $C_{p,Betz}$ can \textcolor{blue}{also be retrieved} as 

\begin{equation}
    C_{p,Betz} = \frac{64C_T'}{(4+C_T')^3}.
    \label{eqn:cp_betz}
\end{equation}

\noindent It is important to note that equations \ref{cp_actuator_disc} and \ref{eqn:ctstar_model} are only strictly valid when $U_{T,i}$ is the same for all turbines. However, in this study we only consider wind farms in which $C_T'$ is the same for all turbines, and in this case, equations \ref{cp_actuator_disc} and \ref{eqn:ctstar_model} can be applied in an approximate manner \textcolor{blue}{even} when $U_{T,i}$ varies throughout the farm.

\par We used the new model of $M$ (equation \ref{eqn:M_model}) with the two-scale momentum theory \textcolor{blue}{(equation \ref{eqn:windfarmmomentum}) with $\gamma = 2$} to predict wind farm power, i.e., solving

\begin{equation}
    \textcolor{blue}{\left(C_T^* \frac{\lambda}{C_{f0}}+1\right) \beta^2 = \frac{1 + \frac{1}{C_{f0}}\frac{H_F}{L}\left(1-\beta^2\right)  - \frac{\tau_{t0}}{\tau_{w0}}}{\beta \left(1-\frac{\tau_{t0}}{\tau_{w0}}\right) }}
    \label{eqn:model_beta}
\end{equation}

\noindent \textcolor{blue}{for $\beta$, which is then substituted into equation \ref{cp_actuator_disc} to calculate $C_p$. The definitions of $C_p$, $C_T^*$ and $C_T'$ are summarised in table \ref{tab:coefficients}.}

\begin{table}
  \begin{center}
\def~{\hphantom{0}}
  \begin{tabular}{c c c p{5cm}}
        Symbol & Name & Formula & Note \\[3pt] \hline
         $C_p$ & Average turbine power coefficient & $\frac{\sum_{i=1}^n P_i}{\frac{1}{2}\rho U_{F0}^3nA}$ & $U_{F0}$ is the undisturbed farm-layer-averaged wind speed. \\ \\
        $C_p^*$ & Average `internal' turbine power coefficient & $\frac{\sum_{i=1}^nP_i}{\frac{1}{2}\rho U_{F}^3nA}$ & $U_{F}$ is the farm-layer-averaged wind speed with the turbines present. \\ \\
        $C_T'$ & Turbine resistance coefficient & $\frac{T_i}{\frac{1}{2}\rho U_{T,i}^2 A}$ & $U_T$ is the wind speed averaged over the rotor swept area. \\ \\
        $C_T^{*}$ & Average `internal' turbine thrust coefficient & $\frac{\sum_{i=1}^n T_i}{\frac{1}{2}\rho U_{F}^2 nA}$ & Determined by turbine operating conditions and layout. \\

  \end{tabular}
  \caption{Summary of turbine power and thrust coefficients.}
  \label{tab:coefficients}
  \end{center}
\end{table}

\par We compared the predictions against finite wind farm LES results from \citet{Wu2017}, \citet{Allaerts2017} and \citet{Lanzilao2022}. We selected these studies because they simulated large wind farms (longer than 15km in the streamwise direction). They published the undisturbed vertical shear stress profile which is a required input to the model of $M$. These studies were also selected because the results are normalised by the power of an isolated turbine rather than the front row turbine power. The 3 studies performed LES of 4 farms with staggered turbine layouts and 6 farms with aligned layouts.

\par \textcolor{blue}{\citet{Kirby2022} reported that $C_T^*$ varied with turbine layout due to turbine-wake interactions. The upper limit of $C_T^*$ was predicted well by equation \ref{eqn:ctstar_model}. Staggered layouts will tend to have $C_T^*$ values close to this upper limit; hence, equations \ref{eqn:ctstar_model} and \ref{eqn:model_beta}  can be used to predict their $C_p$ and give a direct comparison with staggered wind farm LES. \citet{Kirby2022} also showed that turbine wake interactions could reduce $C_T^*$ by up to about 20\% depending on turbine spacings and wind direction. However, the effect of turbine spacing on $C_T^*$ was not studied explicitly for any specific wind direction; therefore, we cannot make a direct comparison with LES of aligned layouts. Nonetheless, we used a 20\% reduced $C_T^*$ value to predict a lower limit to $C_p$ (with the largest turbine-scale loss expected for aligned layouts with a small turbine spacing). This allows a qualitative comparison with LES of aligned wind farms, showing whether our model can correctly capture the observed trends of farm performance under different ABL conditions.}

\par We used the analytical model to predict the farm-averaged power normalised by the isolated turbine power, i.e., $C_p/C_{p,Betz}$. \citet{Allaerts2017} and \citet{Lanzilao2022} used an actuator-disc no rotation model for the turbines with $C_T'=1.33$. For these studies we used $C_T^*$ from the actuator-disc theory (equation \ref{eqn:ctstar_model}) as an upper limit for $C_T^*$ in \ref{eqn:model_beta} and then used the expression $C_p/C_{p,Betz}=\beta^3$. For this $C_T'$ value, the actuator-disc theory gives $C_T^*=0.75$. \citet{Wu2017} used an actuator-disc with rotation implementation for the turbines. \textcolor{blue}{As the value of $C_T'$ was unknown,} we used the published thrust coefficient curves \citep{Wu2015} \textcolor{blue}{to find} an upper limit to $C_T^*$. For the lower limit, we used a 20\% reduced $C_T^*$ value in equation \ref{eqn:model_beta} and then the expression $C_p/C_{p,Betz}=\beta^30.8^{1.5}$ (assuming that $C_T'$ is unaffected by the turbine layout). Note that this expression comes from equation \ref{cp_actuator_disc} using a $C_T^*$ value which is 80\% of the upper limit.

\par \citet{Lanzilao2022} normalised the turbine power by the power of an imaginary turbine row 10km upstream of the farm. However, small reductions in wind speed were observed 10km upstream for the stratified boundary layer. We instead normalised farm powers from this study by the imaginary upstream power for the neutrally stratified case. This is because the neutrally stratified case had a much smaller reduction in wind speed upstream of the farm. The imaginary upstream turbine power of the staggered and aligned farms in the stratified boundary layer were 6.24MW and 6.17MW respectively \citep{Lanzilao_personal}. For the staggered farm in the neutrally stratified case it was 6.61MW \citep{Lanzilao_personal}. 

\par
\begin{figure}
  \centerline{\includegraphics[width=0.75\linewidth]{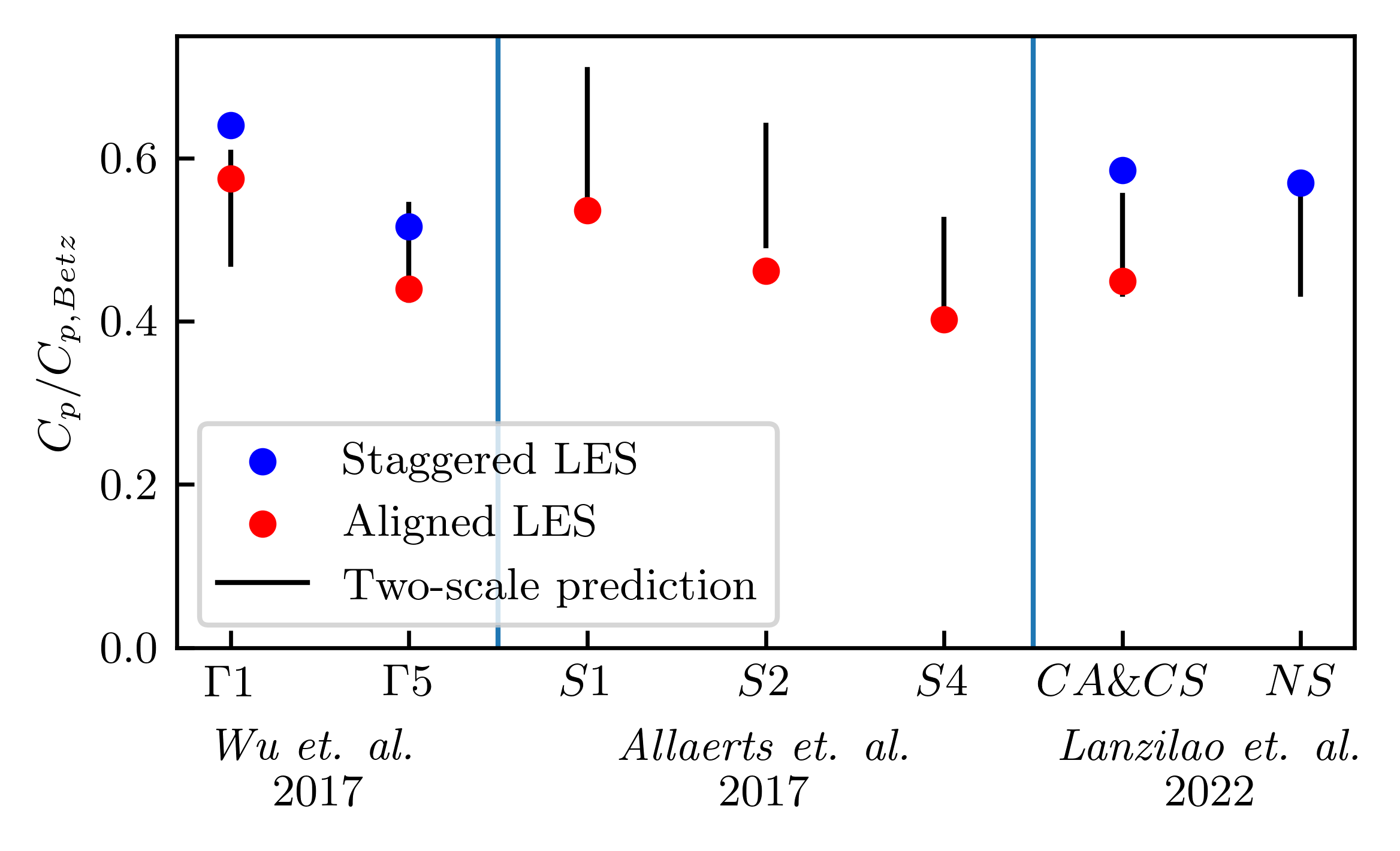}}
  \caption{Comparison of farm-average performance predicted by the two-scale model and results from LES. \textcolor{blue}{The upper end of each black line corresponds to the upper limit of farm performance predicted by the two-scale model (i.e., without 'turbine-scale loss') whereas the lower end corresponds to the lower limit predicted assuming that the turbine-wake interaction causes a 20\% reduction of $C_T^*$.}}
  \label{fig:cp_results}
\end{figure}

 \textcolor{blue}{A summary of comparisons} between the model predictions and LES is shown in figure \ref{fig:cp_results}. Generally, the staggered LES results are close to the upper limits predicted by the two-scale model. The aligned layouts are close to the lower limit of the predictions, except for the results from $\Gamma1$ in \citet{Wu2017} where the LES power is higher than the model predictions. The analytical model predicts well the decrease in farm performance with decreasing \textcolor{blue}{capping} inversion layer height \textcolor{blue}{(1000m, 500m and 250m for cases S1, S2 and S4, respectively)} observed by \citet{Allaerts2017}. For the results of \citet{Lanzilao2022}, the staggered wind farms had very similar performances in the neutral \textcolor{blue}{(NS)} and conventionally neutral \textcolor{blue}{(CS)} boundary layers. The two-scale predictions are exactly the same for these two cases because the shear stress profiles were identical.

\par \textcolor{blue}{Note that we} made direct comparisons \textcolor{blue}{between the two-scale model predictions (without turbine-scale losses) and LES for} staggered layouts because \textcolor{blue}{these layouts are close to `optimal' and give} a $C_T^*$ value close to the \textcolor{blue}{theoretical value given by equation \ref{eqn:ctstar_model} (irrespective of turbine spacing)}. For the aligned layouts, the exact value of $C_T^*$ depends on turbine spacing \textcolor{blue}{and} therefore a direct comparison is not currently possible. Table \ref{tab:cp_results} shows the percentage error in farm power \textcolor{blue}{predicted} for the four staggered farms. When normalised by the farm power \textcolor{blue}{predicted by LES}, the error was typically around 5\% or less. Normalising by the isolated turbine power \textcolor{blue}{predicted theoretically} gave errors of around 3\% or less. 

\begin{table}
  \begin{center}
\def~{\hphantom{0}}
  \begin{tabular}{c|cccc}
        & \multicolumn{2}{c}{\textit{Wu et. al.} 2017}   & \multicolumn{2}{c}{\textit{Lanzilao et. al.} 2022}  \\[3pt]
          & $\Gamma=1K/km$ & $\Gamma=5K/km$ & CS & NS\\
        $\lambda$ & 0.0160 & 0.0160 & 0.0314 & 0.0314\\
        $C_{f0}$ & 0.00557 & 0.00501 & 0.00314 & 0.00314 \\
        $H_{F}/L$ & 0.00875 & 0.00875 & 0.0188 & 0.0188\\
        $\tau_{t0}/\tau_{w0}$ & 0.571 & 0.391 &  0.475 & 0.475\\
        $\beta_{model}$ & 0.847 & 0.816 & 0.821 & 0.821\\
        $M_{model}$ & 2.41 & 2.40 & 5.74 & 5.74\\
        $C_{p,model}/C_{p,Betz}$ & 0.607 & 0.543 & 0.554 & 0.544\\
        $C_{p,LES}/C_{p,Betz}$ & 0.639 & 0.521 & 0.585 & 0.570\\
        $(C_{p,model}-C_{p,LES})/C_{p,LES}$ & -5.00\% & 4.33\%& -5.39\% & -2.87\%\\
        $(C_{p,model}-C_{p,LES})/C_{p,Betz}$ & -3.20\% & 2.25\% & -3.15\% & -1.63\%\\
  \end{tabular}
  \caption{Model parameters and percentage errors in predicting the average power of staggered wind farm LES.}
  \label{tab:cp_results}
  \end{center}
\end{table}

\FloatBarrier

\section{Discussion}\label{discussion}

\FloatBarrier

\par This study proposes a novel framework for modelling \textcolor{blue}{farm-atmosphere interactions}. Equation \ref{eqn:M_decomp} is a general expression for the farm momentum increase as the sum of contributions from different mechanisms. This allows each mechanism to be modelled separately. \textcolor{blue}{In this study we considered contributions from} net momentum advection (including farm blockage), PGF and entrainment. Equation \ref{eqn:M_decomp} could provide a framework for combining different \textcolor{blue}{or improved models of each mechanism} in the future. 

\par We have proposed simple analytical models for each mechanism increasing farm momentum supply. The models make quasi-1D and steady flow assumptions. The simple momentum availability factor model can be coupled with the two-scale momentum theory to predict farm power. The predictions for staggered farms had a typical error of around 5\% or less. The model predictions are for the \textcolor{blue}{average power of a finite-sized farm} rather than just the power in the fully developed region, i.e., the model \textcolor{blue}{captures} the effect of the development region and the farm size. 

\par The only environmental parameter required as an input to the model is the undisturbed vertical shear stress profile of the ABL. The model seems to capture most of the impact of \textcolor{blue}{ABL} height. The proposed model also \textcolor{blue}{considers} the effect of farm \textcolor{blue}{blockage (or the reduction of wind speed upstream of the farm) as part of the advection modelling.} This is a physics-based approach \textcolor{blue}{to instantly predicting the power of a finite-sized farm} without tuning or empirical coefficients.

\par Our model derivation suggests the effects of farm blockage and increased PGF could (at least partly) counteract. \textcolor{blue}{Note that this is a direct result of our assumption that the pressure changes at the front and rear of the farm are of equal magnitude.} A strong free-atmosphere stratification reduces the wind speed in front of the farm \citep{Wu2017,Lanzilao2022}. This reduces the net advection of momentum into the farm. When this occurs, an additional pressure gradient \textcolor{blue}{tends to be} induced across the farm,  \textcolor{blue}{which} increases the momentum supply to the farm. This suggests the negative \textcolor{blue}{effect of farm blockage on the farm power} could be somewhat counteracted by the increased PGF. \textcolor{blue}{The extent to which these effects counteract will likely depend on atmospheric conditions.} For cases NS and CS in \citet{Lanzilao2022}, the atmospheric conditions are kept constant except for the free-atmosphere stratification. The CNBL \textcolor{blue}{case (CS)} showed a much greater farm blockage yet both cases had a similar farm-averaged power, \textcolor{blue}{supporting the above argument}. \textcolor{blue}{The very recent LES study of \citet{Lanzilao2023} also suggests farm blockage often occurs with an increased PGF. However, the degree to which these effects counteract remains unclear.} Hence, the \textcolor{blue}{relationship} between farm blockage and PGF seems important and requires further investigation in the future.

\par
\begin{figure}
  \centerline{\includegraphics[width=0.75\linewidth]{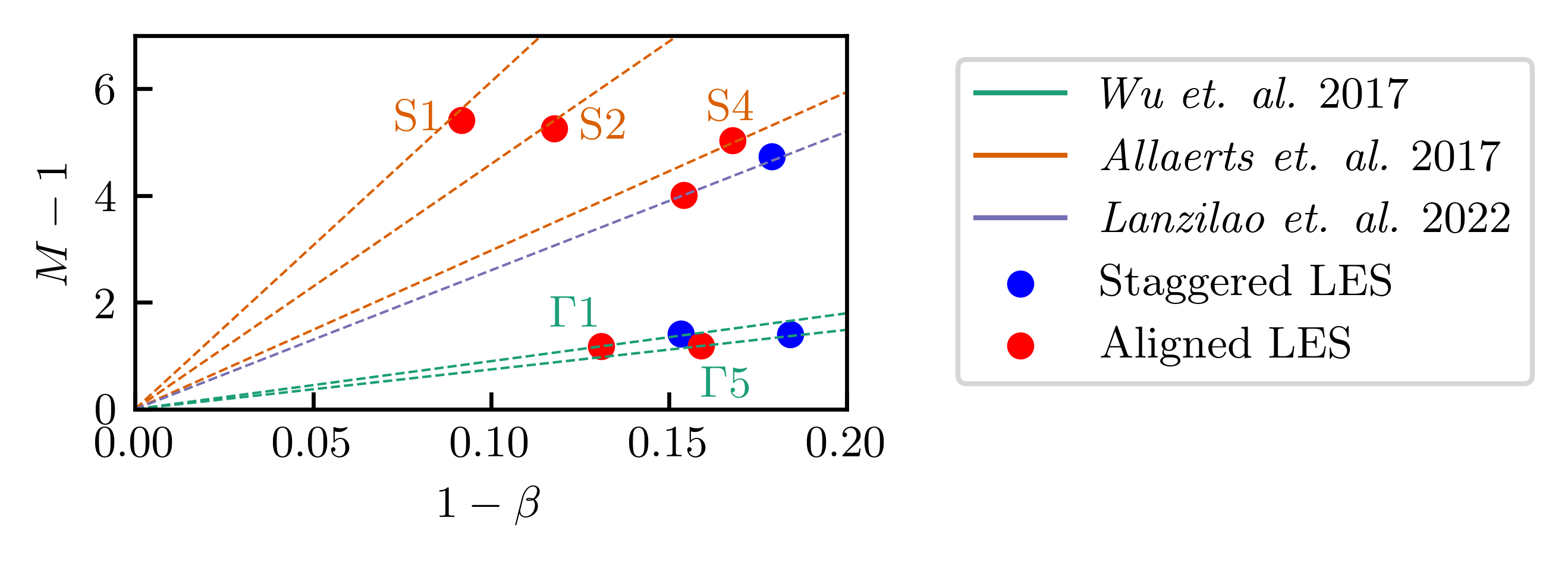}}
  \caption{Predicted values of momentum availability factor ($M$) and farm wind-speed reduction factor ($\beta$). \textcolor{blue}{$\beta$ is calculated using equation \ref{eqn:model_beta} and $M$ using equation \ref{eqn:M_model} for the flow conditions and farm configuration used in the LES. Note that the lines are calculated using equation \ref{eqn:zeta_approx}}.}
  \label{fig:predicted_M}
\end{figure}

 Our model can provide physical insight into how farm efficiency changes with atmospheric conditions. Figure \ref{fig:predicted_M} shows the predicted values of $M$ and $\beta$ for the LES studies considered. The model predicts an approximately linear relationship between $M$ and $\beta$ \textcolor{blue}{for a given atmospheric condition, as observed previously} by \citet{Patel2021}. \textcolor{blue}{The linear relationship is further explained in appendix \ref{section:zeta}.} For the results of \citet{Allaerts2017}, S1 had the highest initial inversion layer and S4 the lowest. Figure \ref{fig:predicted_M} shows that the momentum increase (or the value of $M$) is relatively insensitive to the initial inversion layer height. However, \textcolor{blue}{to achieve a certain momentum increase, a lower inversion layer has a greater wind speed reduction (i.e. a larger value of $(1-\beta)$ in figure \ref{fig:predicted_M}).} The lower wind speed results in S4 having a lower farm efficiency than S1 (see figure \ref{fig:cp_results}).

\par
\begin{figure}
  \centerline{\includegraphics[width=\linewidth]{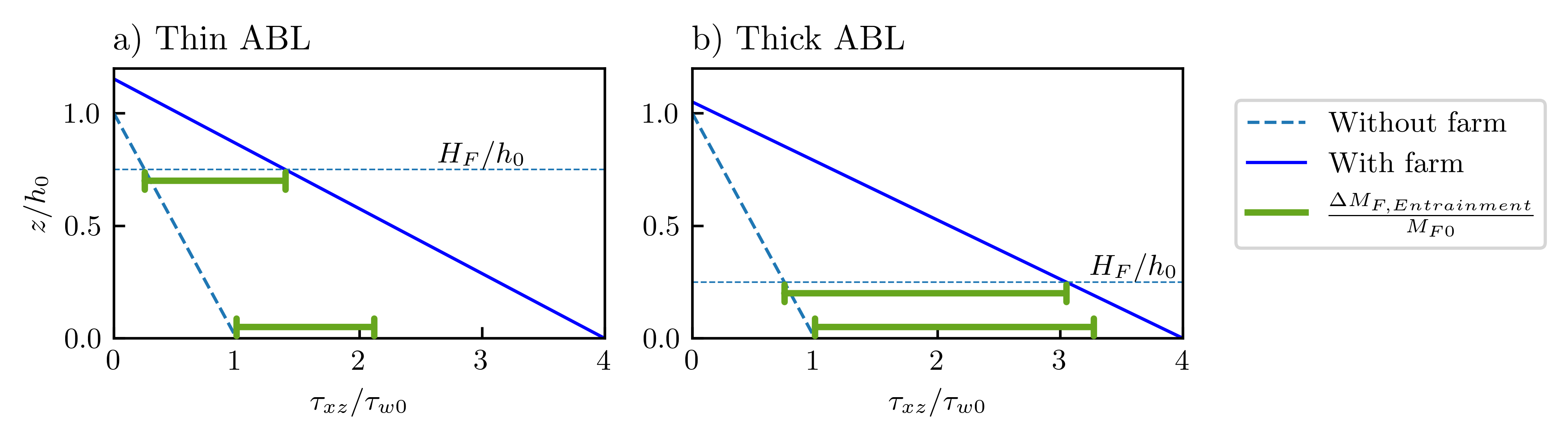}}
  \caption{Contribution of entrainment to farm momentum availability factor $M$ for a) \textcolor{blue}{thin} initial ABL and b) \textcolor{blue}{thick} initial ABL.}
  \label{fig:bl_height_entrain}
\end{figure}

\begin{figure}
  \centerline{\includegraphics[width=0.75\linewidth]{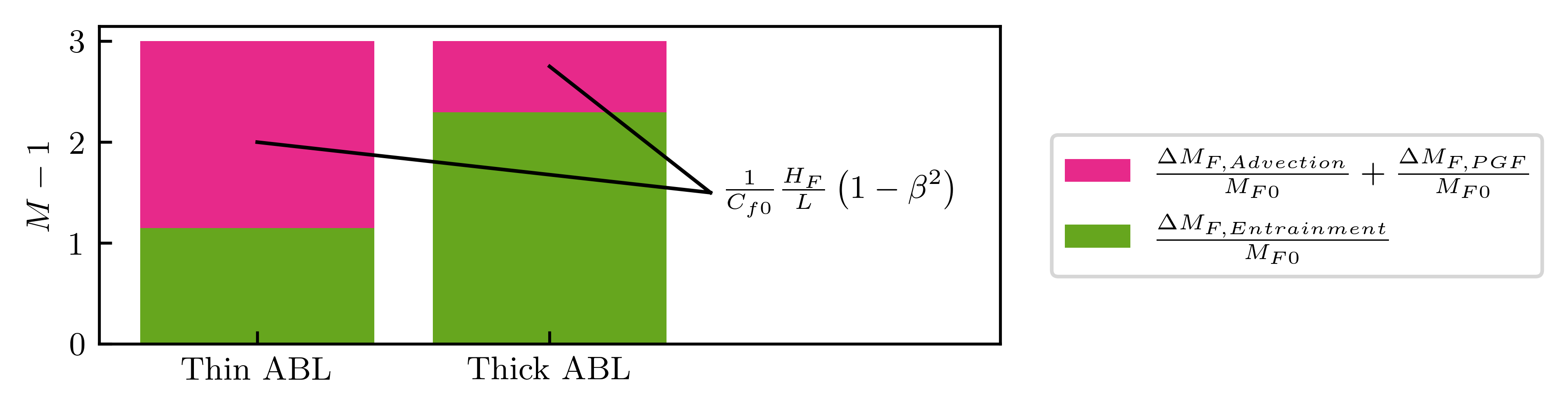}}
  \caption{Decomposition of farm momentum availability factor $M$ into entrainment, and advection and PGF terms for a) a \textcolor{blue}{thin} initial ABL and b) a \textcolor{blue}{thick} initial ABL. Note that the heights of green bars in this figure correspond to the width of green bars in figure \ref{fig:bl_height_entrain}.}
  \label{fig:M_decomp}
\end{figure}

 A \textcolor{blue}{thicker} ABL can increase the farm momentum supply with a smaller wind speed reduction. To illustrate this, we consider the momentum response of two ABLs; a \textcolor{blue}{thin} ABL (with $H_F/h_0=0.75$) and a \textcolor{blue}{thick} ABL (with $H_F/h_0=0.25$). For both ABLs we use a typical value for large offshore farms of $\frac{1}{C_{f0}}\frac{H_F}{L}=7.5$ and set $M=4$ (which physically means that the `internal' conditions of the farm are adjusted to \textcolor{blue}{achieve} $M=4$). The value of $\beta$ can be calculated using equation \ref{eqn:M_model} for both scenarios. The new and undisturbed shear stress profiles are shown in figure \ref{fig:bl_height_entrain}. The \textcolor{blue}{thin} ABL has a smaller entrainment of momentum at the top of the control volume. The top of the control volume is closer to the top of the boundary layer where the velocity shear is lower. Therefore the \textcolor{blue}{thin} ABL is less efficient at entraining momentum into the farm. With less momentum entrainment, more momentum has to be supplied through advection and PGF in order to \textcolor{blue}{achieve} the set level of $M$ (see figure \ref{fig:M_decomp}). These mechanisms increase the momentum supply by reducing the farm-averaged wind speed (equation \ref{eqn:adv_pgf}). In \textcolor{blue}{thicker} ABLs, more momentum is supplied to the farm through entrainment, and the advection and PGF supply less momentum. \textcolor{blue}{Thicker} ABLs can therefore sustain higher wind speeds in the farm and thus have a higher farm efficiency.

\par This study only presents a preliminary comparison with wind farm LES. To fully validate the proposed model, a comparison with a larger set of LES would be required \textcolor{blue}{which will be the focus of future work. A comparison of model predictions with field observations (such as SCADA data) could also be investigated in future studies.} A limitation of the proposed model is that the quasi-1D analysis captures only the first-order effects. However, the model still gave a close agreement with LES results. More detailed flow physics were not considered \textcolor{blue}{explicitly} (e.g. gravity wave excitation) but could be included in future studies. Future work will also aim to predict $C_T^*$ for different turbine layouts and operating conditions. As an example, \citet{legris2023} \textcolor{blue}{and \citet{Kirby2023}} used a wake model to predict $C_T^*$ for different farm designs. Future studies will also focus on the farm-atmosphere interaction under more realistic atmospheric conditions.

\par It should also be noted that, in reality, the assumption of scale separation is not expected to be strictly valid. \textcolor{blue}{The entrainment of momentum into the wind farm could depend on the exact turbine layout and not just the farm-average wind speed reduction. Therefore, to capture higher-order effects, the entrainment component of the model may need some parameters that depend on the internal (turbine-scale) conditions}.

\FloatBarrier

\section{Conclusions}\label{conclusion}

\par In this study, we proposed an analytical model of the momentum availability factor for large wind farms. This is a simple model to instantly predict the impact of farm-scale flows on farm performance. This study considered changes in net advection, PGF and turbulent entrainment as the factors contributing to the momentum availability. We derived the model using a steady quasi-1D control volume analysis and  assuming self-similar vertical shear stress profiles.  The only environmental parameters required as input to the model are the undisturbed vertical shear stress profiles. We used the new model with the two-scale momentum theory \citep{Nishino2020} to predict large farm power production. The model compared well \textcolor{blue}{with} existing LES of finite wind farms in CNBLs. \textcolor{blue}{The model captured the impact of ABL height and farm size on farm performance.}  A direct comparison with LES of staggered farms showed a typical error of 5\% or less.

\par This study also provides a new framework for modelling \textcolor{blue}{farm-atmosphere interactions}. The momentum availability factor is expressed as the sum of contributions from different physical mechanisms. As such, the different mechanisms can be modelled separately. We used first-order analytical models for the different components. Despite the simplicity, the model predicts farm power with close agreement to finite wind farm LES. Only one \textcolor{blue}{algebraic} equation needs to be solved analytically to calculate the farm wind-speed reduction factor $\beta$ and thus predict farm performance. It also provides a physics-based method to predict farm blockage effects and the impact of \textcolor{blue}{ABL height}. Future studies could use more advanced models for the different components of the momentum availability factor. This could provide more accurate predictions of wind farm power production.  

\backsection[Acknowledgements]{The first author (AK) acknowledges the NERC-Oxford Doctoral Training Partnership in Environmental Research (NE/S007474/1) for funding and training. We also thank \textcolor{blue}{Mr.} Luca Lanzilao, Prof. Johan Meyers and Prof. Richard Stevens for providing additional data from their studies.}

\backsection[Funding]{This work was supported by the Natural Environmental Research Council (NERC) award NE/S007474/1.}

\backsection[Declaration of Interests]{The authors report no conflict of interest.}

\backsection[Author ORCID]{A. Kirby, https://orcid.org/0000-0001-8389-1619; T. Nishino, https://orcid.org/0000-0001-6306-7702.}

\backsection[Author contributions]{A.K. and T.N. derived the analytical model. A.K. made a comparison of the model and existing LES results in the literature, and A.K. wrote the paper with corrections from T.N. and T.D.D.}

\appendix

\section{Approximate expression for wind extractability factor $\zeta$}\label{section:zeta}

\FloatBarrier

\par \citet{Patel2021} used twin NWP simulations to calculate $M$ for a realistic wind farm site. They found, for most cases, that $M$ varied almost linearly with the farm induction factor $(1-\beta)$. As such, $M$ can be modelled as

\begin{eqnarray}
M = 1 + \zeta(1-\beta)
\label{eqn:M_linear}
\end{eqnarray}

\noindent where $\zeta$ is the `wind extractability factor' (which is identical to what \citet{Nishino2020} originally proposed as `momentum response factor'). \citet{Kirby2022} reported that $\zeta$ varied with atmospheric conditions and decreased exponentially with wind farm size. The analytical model of $M$ \textcolor{blue}{developed in this study} can be used to derive an approximate expression for $\zeta$, as the equation \ref{eqn:M_model} can be expressed as

\begin{eqnarray}
M &=& \frac{1}{\beta} + \frac{\frac{1}{C_{f0}}\frac{H_F}{L}}{1-\frac{\tau_{t0}}{\tau_{w0}} } \frac{(1-\beta)(1+\beta)}{\beta}.
\end{eqnarray}

\noindent This can than be expressed in terms of farm induction factor $(1-\beta)$, e.g.,

\begin{eqnarray}
M &=& \frac{1}{1- (1-\beta)} + \frac{\frac{1}{C_{f0}}\frac{H_F}{L}}{1-\frac{\tau_{t0}}{\tau_{w0}} } \frac{2(1-\beta)-(1-\beta)^2}{1-(1-\beta)}.
\label{eqn:M_beta_functions}
\end{eqnarray}

\begin{figure}
  \centerline{\includegraphics[width=\linewidth]{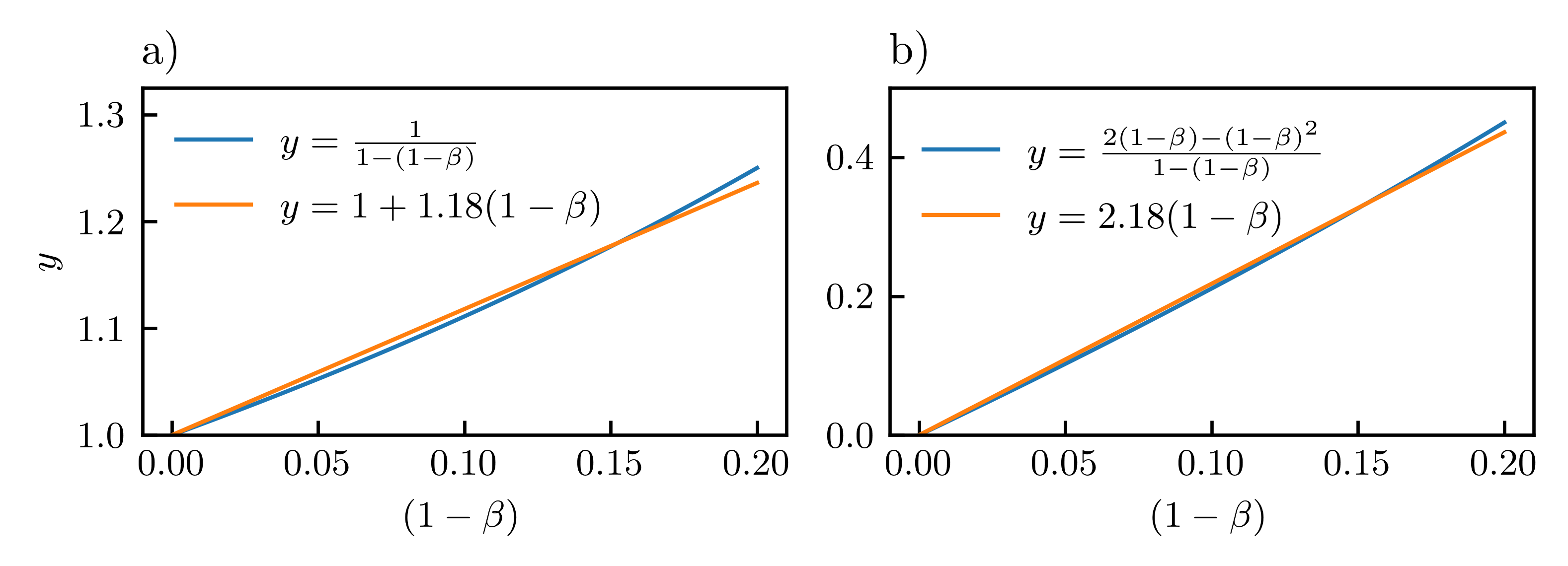}}
  \caption{Linear approximations for terms a) $\frac{1}{1-(1-\beta)}$ and b) $\frac{2(1-\beta)-(1-\beta)^2}{1-(1-\beta)}$ in $M$ expression (equation \ref{eqn:M_beta_functions}).}
  \label{fig:zeta_linear_approx}
\end{figure}

\noindent The functions of $(1-\beta)$ in equation \ref{eqn:M_beta_functions} are approximately linear for a realistic range of $(1-\beta)$ (see figure \ref{fig:zeta_linear_approx}). The analytical model of $M$ (equation \ref{eqn:M_model}) developed in this study therefore predicts the quasi-linear momentum response \textcolor{blue}{observed} in NWP simulations. We use linear interpolation for $(1-\beta)$ between $0$ and $0.2$ (a realistic range) to find linear approximations to the functions of $(1-\beta)$. As such $M$ can be approximated by:

\begin{eqnarray}
M_{approx} &=& 1 + 1.18(1-\beta) + \frac{\frac{2.18}{C_{f0}}\frac{H_F}{L}}{1-\frac{\tau_{t0}}{\tau_{w0}} } (1-\beta).
\label{eqn:M_approx}
\end{eqnarray}

\noindent Equation \ref{eqn:M_approx} can be used to derive an approximate expression for $\zeta$, i.e.,

\begin{eqnarray}
\zeta_{approx} &=& 1.18 + \frac{\frac{2.18}{C_{f0}}\frac{H_F}{L}}{1-\frac{\tau_{t0}}{\tau_{w0}} }
\label{eqn:zeta_approx}
\end{eqnarray}

\noindent using the expression $\zeta = (M-1)/(1-\beta)$. Equation \ref{eqn:zeta_approx} predicts the \textcolor{blue}{inverse relationship} of $\zeta$ with farm size reported by \citet{Kirby2022}. \textcolor{blue}{This expression can be further simplified by assuming that the vertical shear stress profile is linear up to the boundary layer height (as assumed in the discussion of figure \ref{fig:bl_height_entrain} earlier). The expression for $\zeta_{approx}$ then becomes}

\textcolor{blue}{
\begin{eqnarray}
\zeta_{approx} &=& 1.18 + \frac{2.18h_0}{C_{f0}L}
\label{eqn:zeta_approx_h0}
\end{eqnarray}
}

\noindent \textcolor{blue}{where we refer to $L/h_0$ as the farm size ratio.}

\par \textcolor{blue}{The minimum value of $\zeta$ depends on the response of the ABL. If the ABL height is constant (i.e. with a rigid lid), then $\zeta$ tends to zero for an infinitely large farm. The minimum value of $\zeta$ is non-zero even for an infinitely large farm if the ABL height increases in response to the farm. Differentiating equation \ref{eqn:M_model} (with $L=\infty$) gives}

\textcolor{blue}{
\begin{eqnarray}
\zeta = \frac{\textup{d}(M-1)}{\textup{d}(1-\beta)}=\frac{1}{\beta^2}
\end{eqnarray}
}

\noindent \textcolor{blue}{which gives $\zeta=1$ for $(1-\beta)=0$ and $\zeta=1.56$ for $(1-\beta)=0.2$. Note that to derive  equation \ref{eqn:zeta_approx_h0}, we performed a linear regression for $0<(1-\beta)<0.2$, giving a minimum value of $\zeta_{approx}$ of 1.18. Therefore, the minimum value of $\zeta$ for realistic ABLs is expected to be 1.} 

\par
\begin{figure}
  \centerline{\includegraphics[width=\linewidth]{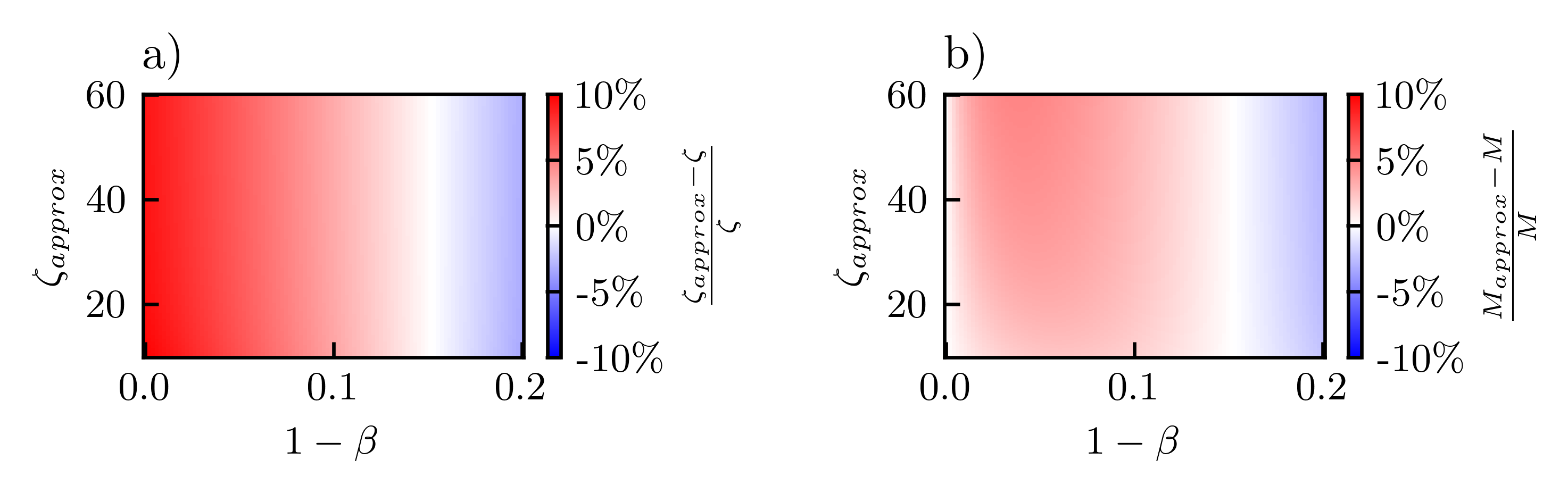}}
  \caption{Percentage error of a) $\zeta_{approx}$ and b) $M_{approx}$.}
  \label{fig:zeta_approx}
\end{figure}

 The discrepancy between $\zeta_{approx}$ and the `true' $\zeta$ (obtained without the linear approximations) is a function only of $\zeta_{approx}$ and $(1-\beta)$. Figure \ref{fig:zeta_approx}a shows the percentage error of $\zeta_{approx}$ for a realistic range of $\zeta_{approx}$ and $(1-\beta)$. The maximum error is approximately 10\% and this occurs for low values of $(1-\beta)$. However, for low values of $(1-\beta)$, the value of $M_{approx}$ is relatively insensitive to the value of $\zeta_{approx}$. Figure \ref{fig:zeta_approx}b shows the percentage error of $M_{approx}$ which is typically less than 5\% for realistic farms.  

\FloatBarrier

\section{\textcolor{blue}{Sensitivity of farm performance to farm length $L$}}

\par
\begin{figure}
  \centerline{\includegraphics[width=\linewidth]{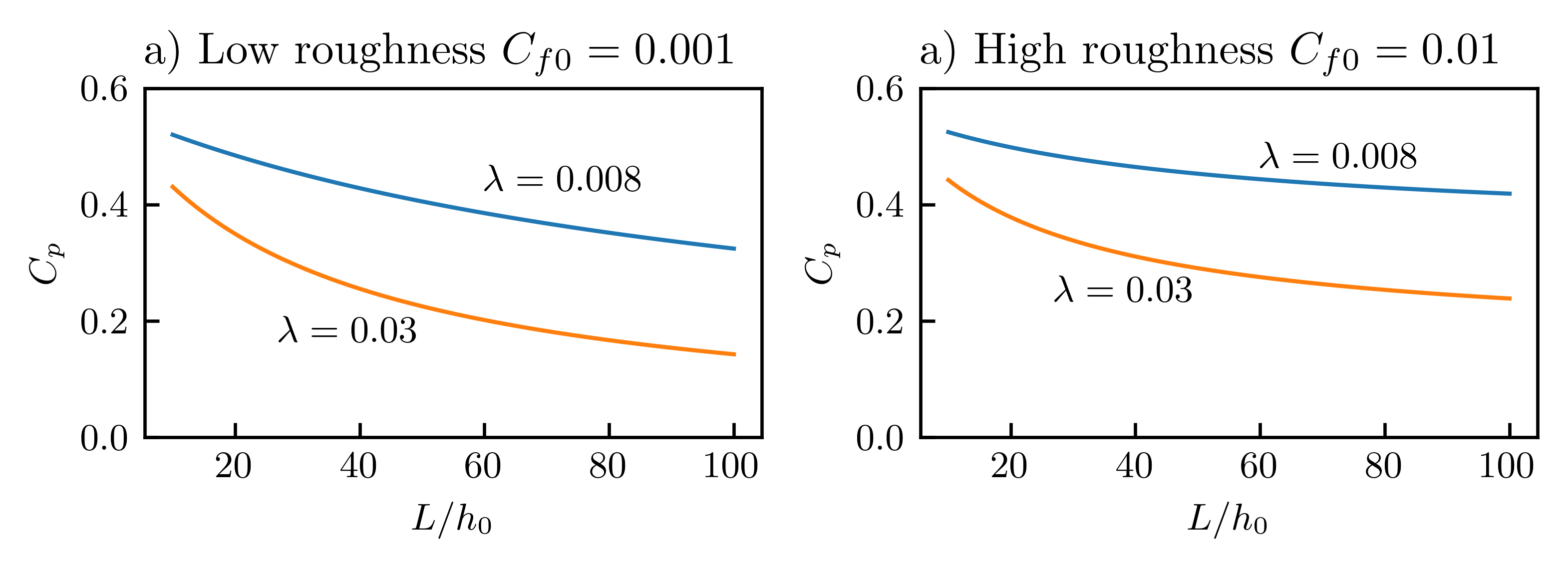}}
  \caption{\textcolor{blue}{Sensitivity of farm-averaged power coefficient $C_p$ to farm size ratio $L/h_0$ for low ($\lambda=0.008$) and high ($\lambda=0.03$) array densities for a) low and b) high surface roughness values, for a fixed turbine resistance coefficient of $C_T' = 1.33$.}}
  \label{fig:farm_length}
\end{figure}

 \textcolor{blue}{Depending on the farm layout, the streamwise farm length $L$ could vary with wind direction. Hence, it is useful to know how the farm power changes with increasing the farm size ratio $L/h_0$. Equations \ref{eqn:windfarmmomentum}, \ref{eqn:M_linear} and \ref{eqn:zeta_approx_h0} are solved for $\beta$, which is then substituted into equation \ref{cp_actuator_disc}. The results are shown for a low surface roughness in figure \ref{fig:farm_length}a and a high roughness in figure \ref{fig:farm_length}b. Note that the results in figure \ref{fig:farm_length} are for a $C_T'$ value of 1.33 and $C_T^*$ calculated using \ref{eqn:ctstar_model}. $L/h_0$ varies from 10 to 100, corresponding to a 10km long farm in a 1km ABL height, up to 30km long farm in a 300m ABL height, for example. Figure \ref{fig:farm_length} shows results for farms with both a high array density ($\lambda=0.03$) and a low density ($\lambda=0.008$). This corresponds to farms with average turbine spacings of $5D$ and $10D$ respectively.}

\par \textcolor{blue}{Figure \ref{fig:farm_length} shows that, for all scenarios, the farm power decreases with increasing the farm length. The farm power is most sensitive to the farm length for small farm size ratios (i.e. shorter wind farms or thicker ABLs). This suggests that as wind farms become larger, the performance could become less sensitive to wind direction. It should also be noted that the farm power is generally higher when the surface roughness is higher, even though the wind extractability factor decreases as the surface roughness increases (equation \ref{eqn:zeta_approx_h0}). This is because the effective array density $\lambda/C_{f0}$ decreases as the surface roughness increases \citep{Nishino2020}.}

\FloatBarrier

\bibliographystyle{jfm}
\bibliography{references.bib}


\end{document}